\documentclass[conference]{IEEEtran}

\usepackage{graphicx}
\usepackage{amsmath,bbm,epsfig,amssymb,amsfonts, amstext, verbatim,amsopn,cite,subfigure,multirow,multicol,lipsum}
\usepackage{balance}
\usepackage{url}
\usepackage{amsfonts}
\usepackage{epsfig}
\usepackage{setspace}
\usepackage{stmaryrd}
\usepackage{psfrag}	
\usepackage{multirow}
\usepackage{float}
\usepackage[process=auto]{pstool}
\usepackage{etoolbox}
\usepackage{algorithm}
\usepackage{algorithmic}
\usepackage{hyperref}

\usepackage{pifont}
\allowdisplaybreaks

\newcommand{\at}[2][]{#1|_{#2}}

\newtheorem{lem}{Lemma}
\newtheorem{remk}{Remark}

\newtoggle{ConfVersion}

\togglefalse{ConfVersion}

\iftoggle{ConfVersion}{%
}{%
}

\EndPreamble
\begin{document}

\title{C-RAN with Hybrid RF/FSO Fronthaul Links:\\
Joint Optimization of  RF Time Allocation and Fronthaul Compression
\vspace{-0.3cm}}
\author{Marzieh Najafi$^{\dag}$, Vahid Jamali$^{\dag}$, Derrick Wing Kwan Ng$^{\ddag}$, and Robert Schober$^{\dag}$ \\
\IEEEauthorblockA{ $^{\dag}$University of Erlangen-Nuremberg, Erlangen, Germany,$\quad$ 
$^{\ddag}$University of New South Wales, Sydney, Australia
}
\vspace{-0.9cm}}

\maketitle

\begin{abstract}
This paper considers the uplink of a cloud radio access network (C-RAN) comprised of several multi-antenna remote radio units (RUs) which send the data that they received from multiple mobile users (MUs) to a central unit (CU) via a wireless fronthaul link. One of the fundamental challenges in implementing C-RAN is the huge data rate required for fronthauling. To address this issue, we employ hybrid radio frequency (RF)/free space optical (FSO) systems for the fronthaul links as they benefit from both the large data rates of FSO links and the reliability of RF links. To efficiently exploit the fronthaul capacity, the RUs employ vector quantization to jointly compress the signals received at their antennas. Moreover, due to the limited available RF spectrum, we assume that the RF multiple-access and fronthaul links employ the same RF resources. Thereby, we propose an adaptive protocol which allocates transmission time to the RF multiple-access and fronthaul links in a time division duplex (TDD) manner and optimizes the quantization noise covariance matrix at each RU such that the sum rate is maximized. Our simulation results reveal that a considerable gain in terms of sum rate can be achieved by the proposed protocol in comparison with benchmark schemes from the literature, especially when the FSO links experience unfavorable atmospheric conditions.
\end{abstract}

\section{Introduction}
Cloud radio access network (C-RAN) is a novel cellular architecture whereby the baseband signal processing is moved from the base stations (BSs) to a cloud-computing based central unit (CU) \cite{Robert_book,C-RAN_Survey_Mao, C-RAN_Survey_VINCENTPOOR}. The BSs operate as remote radio units (RUs) which receive the mobile users' (MUs') data and forward it to the CU via fronthaul links. The CU jointly processes the MUs' data which enables the exploitation of a distributed  multiple-input multiple-output (MIMO) multiplexing gain. Thereby, the main challenge is to convey the signals received at the RUs to the CU via the fronthaul links in an efficient manner as this may require huge data rates, e.g. on the order of Gbits/sec~\cite{ C-RAN_Survey_Mao}. 
Reviews of recent advances in fronthaul-constrained C-RANs are provided in the survey papers \cite{C-RAN_Survey_Mao} and \cite{C-RAN_Survey_VINCENTPOOR}. One popular technique to reduce the data rate requirements of the fronthaul links is to employ compression at the RUs \cite{C-RAN_Wireless_Fronthaul,WeiYu_JSAC,WeiYu_MIMO}. In particular, in~\cite{C-RAN_Wireless_Fronthaul}, the C-RAN uplink was analyzed and an adaptive compression scheme was proposed which minimizes the fronthaul data rate while satisfying a required \textit{block error rate}. Moreover, fronthaul compression optimization for the C-RAN uplink was investigated in~\cite{WeiYu_JSAC} and \cite{WeiYu_MIMO} under a sum-fronthaul capacity constraint and a per-link fronthaul capacity constraint, respectively.

In most of the existing research on C-RANs, fronthaul links are modeled as \textit{abstract} capacity-constrained channels via pure information theoric approaches~\cite{C-RAN_Wireless_Fronthaul,WeiYu_JSAC, WeiYu_MIMO}. Among practical transmission media, optical fiber has been prominently considered as a suitable candidate for fronthaul links mainly due to the large bandwidths available at optical frequencies~\cite{C-RAN_Survey_VINCENTPOOR,Alouini_Backhaul}. However, implementation and maintenance of optical fiber systems are costly. Another competitive candidate technology are free space optical (FSO) systems 
since they provide similar bandwidths as optical fiber systems and are more cost efficient in implementation and maintenance and easy to  upgrade~\cite{Alouini_Backhaul}. 
Unfortunately, the performance of FSO systems significantly deteriorates when the weather conditions are unfavorable, e.g. snowy or foggy  weather~\cite{FSO_Survey_Murat,Alouini_Backhaul}. On the other hand, radio frequency (RF) links are more reliable than FSO links in terms of preserving connectivity but offer lower data rates. Therefore, \textit{hybrid RF/FSO} systems, where RF links are employed to support the FSO links, are appealling candidates for implementation of fronthaul links. These systems complement each other and benefit from both the huge bandwidth of FSO links and the reliability of RF links~\cite{Alouini_Backhaul, MyTCOM}. This motivates us to consider hybrid RF/FSO for the C-RAN fronthaul in this paper. In particular, we consider the \textit{uplink} of a C-RAN with RF multiple-access links and hybrid RF/FSO fronthaul links. Moreover,  due to the limited  RF spectrum, we assume that the RF multiple-access and the fronthaul links share the same RF resources. 

To fully exploit the fronthaul capacity, we employ vector quantization at the RUs, i.e., the signals received at an RU's antennas are jointly quantized in order to exploit the correlation between the signals received at different antennas. Thereby, we formulate a sum-rate maximization problem which optimally divides the time duration between the RF multiple-access and fronthaul links in a time division duplex (TDD) manner and optimizes the  quantization noise covariance matrix at each RU.  Since the optimization problem is non-convex and difficult to solve, we present an equivalent reformulation of the problem. Subsequently, we exploit certain properties of the reformulated problem to develop an efficient algorithm based on golden section search (GSS) and alternating convex optimization (ACO) to obtain a suboptimal solution. Moreover, we note that from the mathematical point of view, the problems considered in~\cite{WeiYu_JSAC} and \cite{WeiYu_MIMO} are special cases of the problem considered in this paper. Unlike~\cite{WeiYu_JSAC} and \cite{WeiYu_MIMO}, in this paper, hybrid RF/FSO fronthaul links are considered and the RF transmission time is optimized and adaptively shared  between the multiple-access and fronthaul links, which to the best of the authors' knowledge, has not been considered in the literature yet. Our simulation results unveil the gains achieved by the proposed RF time allocation and fronthaul compression policies and show that in comparison with an FSO-only fronthaul, the considered hybrid RF/FSO fronthaul ensures a certain minimum achievable sum rate even if the FSO links experience unfavorable atmospheric conditions.


\textit{ Notation:} Boldface lower-case and upper-case letters denote vectors and matrices, respectively. Calligraphic letters are used to denote sets. The superscripts $(\cdot)^{\mathsf{T}}$, $(\cdot)^{\mathsf{H}}$, and $(\cdot)^{-1}$ denote the transpose, Hermitian transpose, and matrix inverse operators, respectively; $\mathbbmss{E}\{\cdot\}$,  $\mathrm{Tr}(\cdot)$, and $|\cdot|$ denote the expectation, matrix trace, and matrix determinant operators, respectively. $\mathbf{A}\succeq \mathbf{0}$ indicates that matrix $\mathbf{A}$ is positive semidefinite;  $\mathbf{I}_n$ represents the $n$-dimentional identity matrix. Moreover, $\mathbb{R}^+$, $\mathbb{R}$, and $\mathbb{C}$ denote the sets of positive real, real, and complex numbers, respectively. We use  $\mathrm{diag}\{\mathbf{A}_1,\dots,\mathbf{A}_n\}$ to denote a block diagonal matrix formed by matrices $\mathbf{A}_1,\dots,\mathbf{A}_n$ on the diagonal. Moreover, $[x]^+$ is defined as $[x]^+\triangleq\mathrm{max}\{0,x\}$; $\ln(\cdot)$ denotes the natural logarithm; and $\mathbf{a}\sim\mathcal{CN}(\boldsymbol{\mu},\boldsymbol{\Sigma})$ is used to indicate that $\mathbf{a}$ is a random complex Gaussian vector  with mean vector $\boldsymbol{\mu}$ and covariance matrix $\boldsymbol{\Sigma}$.

\section{System and Channel Models}\label{SysMod}
In this section, we present the system model and the channel models for the multiple-access and fronthaul links.

\subsection{System Model}

\begin{figure}
\centering
\scalebox{0.54}{
\pstool[width=1.8\linewidth]{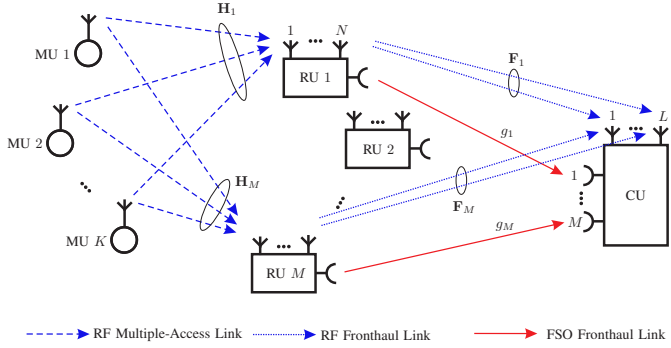}{
\psfrag{U1}[c][c][1]{MU~$1$}
\psfrag{U2}[c][c][1]{MU~$2$}
\psfrag{UK}[c][c][1]{MU~$K$}
\psfrag{H1}[c][c][1]{$\mathbf{H}_1$}
\psfrag{HM}[c][c][1]{$\mathbf{H}_M$}
\psfrag{R1}[c][c][1]{RU~$1$}
\psfrag{R2}[c][c][1]{RU~$2$}
\psfrag{RM}[c][c][1]{RU~$M$}
\psfrag{CU}[c][c][1]{CU}
\psfrag{1}[c][c][1]{$1$}
\psfrag{L}[c][c][1]{$L$}
\psfrag{M}[c][c][1]{$M$}
\psfrag{N}[c][c][1]{$N$}
\psfrag{F1}[c][c][1]{$\mathbf{F}_1$}
\psfrag{FM}[c][c][1]{$\mathbf{F}_M$}
\psfrag{g1}[c][c][1]{$g_1$}
\psfrag{gM}[c][c][1]{$g_M$}
\psfrag{Rf1}[l][c][1]{\hspace{-0.2cm}RF Multiple-Access Link}
\psfrag{Rf2}[l][c][1]{\hspace{-0.2cm}RF Fronthaul Link}
\psfrag{FSO}[l][c][1]{\hspace{-0.2cm}FSO Fronthaul Link}
}}
\caption{C-RAN with hybrid RF/FSO fronthaul links.}
\label{FigSysMod}\vspace{-0.3cm}
\end{figure}
We consider the uplink of a C-RAN where $K$ MUs denoted by MU~$k,\,\, k \in \mathcal{K}=\{1,\dots,K\}$, communicate with a CU via $M$ intermediate RUs denoted by RU~$m,\,\,m\in \mathcal{M}=\{1,\dots,M\}$. Fig.~\ref{FigSysMod} schematically shows the considered communication setup. We assume that the RUs and CU are fixed nodes whereas the MUs can be mobile nodes. Moreover, because of the large distance, the direct link between the MUs and the CU is not exploited. There are two transmission links in our system model: $i$) the MU-RU RF multiple-access link and $ii$) the RU-CU hybrid RF/FSO fronthaul links. Each MU is equipped with a single RF antenna whereas each RU has $N $ RF antennas as well as one aperture FSO transmitter pointed towards the CU. The CU is equipped with $M$ photodetectors, each directed to its corresponding RU, as well as $L$ RF antennas. We assume that the photodetectors are sufficiently spaced such that interference between FSO links is avoided\footnote{The minimum spacing between photodetectors required to avoid cross talk mainly depends on the divergence angle of the FSO beams, the distance between the RUs and the CU, and the relative position of the RUs~\cite{FSO_Survey_Murat}.}. Furthermore, we assume block fading, i.e., the fading coefficients are constant during one fading block but may change from one fading block to the next.
Throughout this paper, we assume that the CU has the instantaneous CSI of all RF and  FSO links and is responsible for determining the transmission strategy and informing it to all nodes. Moreover, we assume that the channel states change slowly enough such that the signaling overhead caused by channel estimation and feedback is negligible compared to the amount of information transmitted in one fading block.

\vspace{-0.16cm}
\subsection{Channel Model} 
 In the following, we describe the channel models for the multiple-access and fronthaul links.

\subsubsection{Multiple-Access Link}
We assume a standard additive white Gaussian noise (AWGN) channel with fading for the RF multiple-access links. All MUs transmit simultaneously using the same frequency band. The  signal received at RU~$m$ is denoted by $\mathbf{y}_m\in \mathbb{C}^{N \times 1}$ and is given by
\begin{IEEEeqnarray}{lll}\label{Eq:Signal_RF1}
\mathbf{y}_m=\mathbf{H}_m\mathbf{x}+\mathbf{n}_m,\,\,\forall m\in\mathcal{M},
\end{IEEEeqnarray}
where  $\mathbf{x}\in \mathbb{C}^{K \times 1}$ is the vector of signals transmitted by the $K$ MUs. We assume $\mathbbmss{E}\{\mathbf{x}\mathbf{x}^{\mathsf{H}}\}=\mathrm{diag}\{P_1,P_2,\dots,P_K\}\triangleq \boldsymbol{\Sigma}$, i.e., the signals transmitted by different MUs are independent; and $P_k$ is the transmit power of MU~$k$. In addition, $\mathbf{n}_m\in \mathbb{C}^{N \times 1}$ is the noise vector at  RU~$m$. The elements of $\mathbf{n}_m$, i.e., the noise at each antenna, are modelled as mutually independent zero-mean complex AWGN with variance $\sigma^2$. Moreover, $\mathbf{H}_m \in \mathbb{C}^{N \times K}$ denotes the channel matrix corresponding to the RF multiple-access link from the MUs to RU~$m$.

\subsubsection{Fronthaul Links}
The fronthaul links are hybrid RF/FSO. For the FSO links, the aperture transmitter of each RU is directed to the corresponding photodetector at the CU. We assume an intensity modulation direct detection (IM/DD) FSO system with on-off keying (OOK) modulation. 
Particularly, after removing the ambient background light intensity, the signal intensity detected at the $m$-th photodetector of the CU is denoted by $\tilde{{y}}_m\in \mathbb{R}$ and modelled as \cite{FSO_Vahid}
\begin{IEEEeqnarray}{lll}\label{Eq:Signal_FSO}
\tilde{y}_m={g}_m\tilde{x}_m+\tilde{n}_m,\,\,\forall m\in\mathcal{M},
\end{IEEEeqnarray}
where $\tilde{x}_m\in \{0,\tilde{P}_m\}$ is the OOK-modulated symbol at RU~$m$ and $\tilde{n}_m\in \mathbb{R}$ is the zero-mean real-valued AWGN shot noise with variance $\delta^2$ at the CU caused by ambient light. Moreover, $\tilde{P}_m$ is the  maximum allowable transmit power of RU~$m$ over the FSO link which is mainly determined by eye safety regulations \cite{FSO_Survey_Murat}. Furthermore, ${g}_m\in \mathbb{R}^+$ denotes the FSO channel gain from RU~$m$ to the CU's $m$-th photodetector. The capacity of the FSO link from RU~$m$ to the CU for OOK signaling is given by~\cite{FSO_Vahid}
\begin{IEEEeqnarray}{lll}\label{Eq:FSO_C}
C_m^{\mathrm{fso}}=W^{\mathrm{fso}}\Bigg[1-\dfrac{1}{\sqrt{2\pi}}  \displaystyle\int\limits_{ - \infty }^\infty e^ {-t^2} \mathrm{log}_2 \Bigg\{ 1+e^{- \dfrac{\tilde{P}_m^2 g_m^2}{2\delta^2}}\nonumber \\
\times\Bigg[ e^{ \dfrac{2t\tilde{P}_m g_m}{\sqrt{2\delta^2}}} +e^{ - \dfrac{2t\tilde{P}_m g_m}{\sqrt{2\delta^2}}} +e^{ - \dfrac{\tilde{P}_m^2g_m^2}{2\delta^2}} \Bigg]  \Bigg\} \mathrm{d}t\Bigg]\,\text{bits/sec}, \quad
\end{IEEEeqnarray}
where $W^{\mathrm{fso}}$ is the bandwidth of the FSO signal.

For simplicity of implementation, we employ a TDD protocol to ensure that the RF fronthaul links are orthogonal with respect to (w.r.t.) each other and also orthogonal to the RF multiple-access link. The RF signal of RU~$m$ received at the CU is denoted by $\bar{\mathbf{y}}_m\in \mathbb{C}^{L \times 1}$ and given by
\begin{IEEEeqnarray}{lll}\label{Eq:Signal_RF2}
\bar{\mathbf{y}}_m=\mathbf{F}_m\bar{\mathbf{x}}_m+\bar{\mathbf{n}}_m,\,\,\forall m\in\mathcal{M},
\end{IEEEeqnarray}
where  $\bar{\mathbf{x}}_m\in \mathbb{C}^{N \times 1}$ is the vector of signals transmitted over the antennas of RU~$m$. We assume that $\mathbbmss{E}\{\bar{\mathbf{x}}^{\mathsf{H}}_m\bar{\mathbf{x}}_m\}=\bar{P}_m$ holds, where $\bar{P}_m$ is the fixed transmit power of RU~$m$ over the RF fronthaul links, and $\bar{\mathbf{n}}_m\in \mathbb{C}^{L \times 1}$ denotes the noise vector at the CU. The noise at each antenna of the CU, i.e., each element of $\bar{\mathbf{n}}_m$, is modelled as zero-mean complex AWGN with variance $\sigma^2$. Moreover, $\mathbf{F}_m\in \mathbb{C}^{L \times N}$ denotes the channel matrix corresponding to the RF fronthaul link from RU~$m$ to the CU. The capacity of the RF fronthaul link between RU~$m$ and the CU is obtained via water filling as \cite{FSO_Vahid}
\begin{IEEEeqnarray}{lll}\label{Eq:RF_C}
C_{m}^{\mathrm{rf}}=W^{\mathrm{rf}}\sum_{j=1}^{\min\{N,L\}}\left[\mathrm{log_2}\left\{\dfrac{\mu\chi_{m,j}^2}{\sigma^2}\right\}\right]^+\,\text{bits/sec}, \quad
\end{IEEEeqnarray}
 where  $W^{\mathrm{rf}}$ is the bandwidth of the  RF signal. In (\ref{Eq:RF_C}), $\chi_{m,j}$ is the $j$-th singular value of $\mathbf{F}_m$ and $\mu$ is the water level which is chosen to satisfy the power constraint as the solution of the following equation
\begin{IEEEeqnarray}{lll}\label{Eq:mu}
\sum_{j=1}^{\min\{N,L\}}\left[\mu-\dfrac{\sigma^2}{\chi_{m,j}^2}\right]^+=\bar{P}_m.
\end{IEEEeqnarray}

\section{Problem Formulation}

In this section, we propose an adaptive protocol for RF time allocation and fronthaul compression and formulate a sum-rate maximization problem for optimization of the protocol.

\subsection{Fronthaul RF Time Allocation}
\begin{figure}
\centering
\scalebox{0.49}{
\pstool[width=2\linewidth]{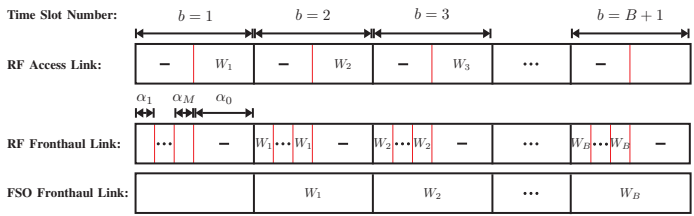}{
\psfrag{Rf1}[l][l][1]{\textbf{RF Access Link:}}
\psfrag{Rf2}[l][l][1]{\textbf{RF Fronthaul Link:}}
\psfrag{Fso}[l][l][1]{\textbf{FSO Fronthaul Link:}}
\psfrag{Bnum}[l][l][1]{\textbf{Time Slot Number:}}
\psfrag{b1}[c][c][1.2]{$b=1$}
\psfrag{b2}[c][c][1.2]{$b=2$}
\psfrag{b3}[c][c][1.2]{$b=3$}
\psfrag{bb}[c][c][1.2]{$b=B+1$}
\psfrag{W1}[c][c][1]{$W_1$}
\psfrag{W2}[c][c][1]{$W_2$}
\psfrag{W3}[c][c][1]{$W_3$}
\psfrag{WB}[c][c][1]{$W_{B}$}
\psfrag{a1}[c][c][1.2]{$\alpha_1$}
\psfrag{a2}[c][c][1.2]{$\alpha_M$}
\psfrag{am}[c][c][1.2]{$\alpha_0$}
}}
\caption{Proposed transmission protocol for C-RAN with hybrid RF/FSO fronthaul link where $W_b$ is the data transmitted by the MUs in time~slot~$b$.}
\label{FigBlock}\vspace{-0.3cm}
\end{figure}

Due to the limited available RF spectrum, we assume that the multiple-access and fronthaul RF links utilize the same RF resource. Hence, the advantages of employing a hybrid RF/FSO system for the fronthaul link come at the expense of a bandwidth reduction for the multiple-access link. In the following, we propose an adaptive protocol which allocates the RF time between the multiple-access and the fronthaul links in a TDD manner\footnote{Although RF time sharing between multiple-access and fronthaul links based on TDD is in general suboptimal, we adopt it here as this leads to a simple implementation at the transceivers since interlink interference is avoided.}. To formally present our protocol, we assume that each transmission block, i.e., one fading block, is divided into $B+1$ time slots indexed by $b\in\{1,\dots,B+1\}$. Hereby, the RUs transmit the data received from the MUs in time slot $b\in\{1,\dots,B\}$ to the CU in the subsequent time slot $b+1$. We note that in the first time slot, no data is available at the RUs to be sent over the fronthaul links. In addition, in the last time slot, $B+1$, the MUs do not transmit any new data. This leads to a rate reduction by a factor of $\frac{B}{B+1}$. However, as $B\to \infty$, we have $\frac{B}{B+1}\to 1 $. Moreover, let $\alpha_m \in [0,1]$ denote the fraction of RF time which is allocated to the fronthaul link to forward the data of RU~$m$ to the CU. Hence, a fraction of $\alpha_0 = 1-\sum_{\forall m}\alpha_{m}$ of the RF time is available for the multiple-access link. The considered transmission protocol is schematically illustrated in Fig.~\ref{FigBlock}. For future reference, we define $\boldsymbol{\alpha}=[\alpha_0,\alpha_1,\dots,\alpha_M]^\mathsf{T}\in\mathcal{A}$ where $\mathcal{A}=\big\{\boldsymbol{\alpha}|\sum_{m=0}^{M}\alpha_m=1,\,\alpha_m\in[0,1],\,\, \forall m \in \{0,1,\dots,M\}\big\}$.

\subsection{Fronthaul Compression}
We assume that each RU employs compress-and-forward and quantizes the received signal $\mathbf{y}_m$ into $\hat{\mathbf{y}}_m,\,\,m\in\mathcal{M} $. In particular, RU~$m$ quantizes the received imphase/quadrature (I/Q) samples with a sampling rate of $f_s\geq W^{\mathrm{rf}} $ and forwards the compressed signals to the CU through hybrid RF/FSO fronthaul links. According to rate-distortion theory, the Gaussian quantization test channel which relates $\mathbf{y}_m$ to $\hat{\mathbf{y}}_m$ is given by~\cite{WeiYu_JSAC}
\begin{IEEEeqnarray}{lll}\label{Eq:noise_Q_global}
\hat{\mathbf{y}}_m=\mathbf{y}_m+\mathbf{z}_m,
\end{IEEEeqnarray}
where $\mathbf{z}_m\in \mathbb{C}^{N\times 1}\sim \mathcal{CN}(\mathbf{0}_{N}, \mathbf{D}_m)$ is the quantization noise and $\mathbf{D}_m=\mathbbmss{E}\{\mathbf{z}_m\mathbf{z}_m^{\mathsf{H}}\}$ is the distortion matrix at RU~$m$. The mean square distortion between the I/Q sample vector $\mathbf{y}_m$  and the corresponding quantized vector $\hat{\mathbf{y}}_m$ is given by  the main diagonal entries of $\mathbf{D}_m$. We consider vector quantization, which exploits the correlation between the received signals at an RU's antennas. Hence, $\mathbf{D}_m$ is  a non-diagonal matrix in general.

\subsection{Sum-Rate Maximization Problem}
The considered sum-rate maximization problem under fronthaul capacity constraint is defined as
\begin{IEEEeqnarray}{cclll}\label{Eq:Local}
&\underset{\boldsymbol{\alpha}\in \mathcal{A},\mathbf{D}_m\succeq \mathbf{0},\,\forall m}{\mathrm{maximize}}\,\,& C_\mathrm{{sum}}=\alpha_0 W^{\mathrm{rf}} I\left(  \mathbf{x},\hat{\mathbf{y}}\right), \\
& \mathrm{subject\,\, to}  &  \alpha_0 f_s I\left(  \mathbf{y}_m,\hat{\mathbf{y}}_m\right)\leq {C}_m^{\mathrm{fso}}+{\alpha}_m{C}_m^{\mathrm{rf}},\,\,\forall m\in\mathcal{M}, \nonumber
  \end{IEEEeqnarray}
where $\mathbf{y}=\left[\mathbf{y}_1^{\mathsf{T}},\mathbf{y}_2^{\mathsf{T}},\dots,\mathbf{y}_M^{\mathsf{T}}\right]^{\mathsf{T}}$ and $\hat{\mathbf{y}}=\left[\hat{\mathbf{y}}_1^{\mathsf{T}},\hat{\mathbf{y}}_2^{\mathsf{T}},\dots,\hat{\mathbf{y}}_M^{\mathsf{T}}\right]^{\mathsf{T}}$. Moreover, for the Gaussian RF multiple-access channel in (\ref{Eq:Signal_RF1}) and the Gaussian quantization test channel in (\ref{Eq:noise_Q_global}), $I\left(\mathbf{x},\hat{\mathbf{y}}\right)$ and $I\left(\mathbf{y}_m,\hat{\mathbf{y}}_m\right)$ are given by \cite{WeiYu_JSAC}
\begin{IEEEeqnarray}{lll}
I\left(  \mathbf{x},\hat{\mathbf{y}}\right)&=   \mathrm{log_2}\dfrac{\left\vert \mathbf{H}\boldsymbol{\Sigma}\mathbf{H}^{\mathsf{H}}+\mathbf{D}+\sigma^2\mathbf{I}_{MN}\right\vert}{\left\vert \mathbf{D}+\sigma^2\mathbf{I}_{MN}\right\vert}\quad \text{and}\label{Eq:C_sum_Local} \\
I\left(  \mathbf{y}_m,\hat{\mathbf{y}}_m\right) &=   \mathrm{log_2}\dfrac{\left\vert \mathbf{H}_m\boldsymbol{\Sigma}\mathbf{H}_m^{\mathsf{H}}+\mathbf{D}_m+\sigma^2\mathbf{I}_{N}\right\vert}{\left\vert \mathbf{D}_m\right\vert},\qquad\label{Eq:Constraint_Local}
\end{IEEEeqnarray}
respectively, where $\mathbf{D}=\mathrm{diag}\{\mathbf{D}_1,\mathbf{D}_2,\dots,\mathbf{D}_M\}$ comprises the distortion matrices of all RUs and $\mathbf{H}=\left[\mathbf{H}_1^{\mathsf{T}},\mathbf{H}_2^{\mathsf{T}},\dots,\mathbf{H}_M^{\mathsf{T}}\right]^{\mathsf{T}}$ is the channel matrix between the MUs and the RUs. 

The problem in (\ref{Eq:Local}) readily reveals the tradeoff which exists for the RF time allocation between the multiple-access and fronthaul links. In particular, an intuitive observation here is that when the quality of the FSO links is sufficiently good allowing arbitrary small distortions, i.e., ${C}_m^{\mathrm{fso}}\to \infty,\, \forall m$, there is no need for an RF fronthaul link and the optimal RF time allocation policy is to simply allocate the available RF time to the multiple-access link, i.e., the optimal RF time allocation is $\alpha_0^*=1$ and $\alpha_m^*=0,\, \forall m\in \mathcal{M}$. On the other hand, when the quality of the FSO links is poor due to e.g. adverse atmospheric conditions, i.e., ${C}_m^{\mathrm{fso}}\to 0,\, \forall m$, then the backup RF fronthaul links are needed, i.e., $\exists \alpha_m> 0,\forall m \in \mathcal{M}$, to ensure a minimum non-zero achievable sum rate.


\section{Proposed RF Time Allocation and Fronthaul Compression Policies}
In this section, we derive a suboptimal solution to optimization problem~(\ref{Eq:Local}) which provides an efficient joint RF time allocation and fronthaul compression policy.

\subsection{Reformulation of the Optimization Problem}
We note that optimization problem (\ref{Eq:Local}) is jointly non-convex in $(\boldsymbol{\alpha},\mathbf{D})$. Hence, finding the globally optimal solution requires large computational complexity. To cope with this issue, we first present a reformulation of the constraints of problem~(\ref{Eq:Local}). Using this reformulation, we propose to employ GSS to find the optimal $\boldsymbol{\alpha}^*$ assuming that the optimal $\mathbf{D}^*$ is known. Subsequently, we tackle the problem of finding $\mathbf{D}^*$ for a given $\boldsymbol{\alpha}$. Since this problem is still non-convex in optimization variable $\mathbf{D}$, we  present a reformulation of the objective function of~(\ref{Eq:Local}) which enables us to derive a suboptimal solution using ACO. Finally, based on these results, we propose an algorithm in Section IV-B which finds an efficient suboptimal solution of (\ref{Eq:Local}).

\subsubsection{Reformulation of the Constraints} In the following lemma, we provide a useful equivalent representation of the constraints in~(\ref{Eq:Local}) to handle $\boldsymbol{\alpha}$.

\begin{lem}\label{LemmaA}
The constraints in (\ref{Eq:Local}) can be written in the following equivalent form such that ${\alpha}_m, \forall m\in \mathcal{M},$ does not explicitly appear in the new constraints:
\begin{IEEEeqnarray}{lll}\label{Eq:E_Constraint}
\mathrm{C}_{\mathcal{S}}:\quad \alpha_0 f_s \sum_{\forall m\in \mathcal{S}}G_m(\mathcal{S})I\left(\mathbf{y}_m,\hat{\mathbf{y}}_m\right)\nonumber\\
\qquad\quad \leq (1-\alpha_0)G(\mathcal{S})+\sum_{\forall m\in \mathcal{S}}G_m(\mathcal{S}){C}_m^{\mathrm{fso}},\,\, \forall\mathcal{S} \subseteq \mathcal{M},\quad\,\,
\end{IEEEeqnarray}
where $G_m(\mathcal{S})=\dfrac{\prod_{\forall m' \in \mathcal{S}}C_{m'}^{\mathrm{rf}}}{C_m^{\mathrm{rf}}}$ , $G(\mathcal{S})=\prod_{\forall m\in \mathcal{S}}{C}_m^{\mathrm{rf}}$, and $\mathcal{S}$ denotes a non-empty subset of $\mathcal{M}$.
\end{lem} 
\begin{IEEEproof}
The proof is provided in the Appendix.
\end{IEEEproof}
The advantage of Lemma~\ref{LemmaA} is that the $(M+1)$-dimensional optimization variable $\boldsymbol{\alpha}$ reduces to the one-dimensional optimization variable $\alpha_0$ at the expense of increasing the number of constraints from $M$ in (\ref{Eq:Local}) to $2^{M}-1$ in (\ref{Eq:E_Constraint}).
To illustrate Lemma~\ref{LemmaA}, let us consider the special case $\mathcal{M}=\{1,2\}$, where the two constraints in (\ref{Eq:Local}) and the equivalent  three constraints in  (\ref{Eq:E_Constraint}) are given by
\begin{IEEEeqnarray}{lll}\label{Eq:Example}
\begin{cases}
\alpha_0 f_s I\left(\mathbf{y}_1,\hat{\mathbf{y}}_1\right) \leq \alpha_1 C_1^{\mathrm{rf}}+{C}_1^{\mathrm{fso}}, \\
\alpha_0 f_s I\left(\mathbf{y}_2,\hat{\mathbf{y}}_2\right) \leq \alpha_2 C_2^{\mathrm{rf}}+{C}_2^{\mathrm{fso}},
 \end{cases}\IEEEyesnumber\IEEEyessubnumber\\
 \begin{cases}
\alpha_0 f_s I\left(\mathbf{y}_1,\hat{\mathbf{y}}_1\right) \leq (1-\alpha_0)C_1^{\mathrm{rf}}+{C}_1^{\mathrm{fso}}, \\
\alpha_0 f_s I\left(\mathbf{y}_2,\hat{\mathbf{y}}_2\right) \leq (1-\alpha_0)C_2^{\mathrm{rf}}+{C}_2^{\mathrm{fso}},\\
\alpha_0 f_s \left(C_2^{\mathrm{rf}}I\left(\mathbf{y}_1,\hat{\mathbf{y}}_1\right)+C_1^{\mathrm{rf}}I\left(\mathbf{y}_2,\hat{\mathbf{y}}_2\right)\right) \\
 \qquad \qquad\qquad\,\, \leq (1-\alpha_0)C_1^{\mathrm{rf}}C_2^{\mathrm{rf}}+C_2^{\mathrm{rf}}{C}_1^{\mathrm{fso}}+C_1^{\mathrm{rf}}{C}_2^{\mathrm{fso}},
 \end{cases} \IEEEyessubnumber
\end{IEEEeqnarray}
respectively. 

\begin{remk}
The equivalence of the constraints in  (\ref{Eq:E_Constraint}) and (\ref{Eq:Local}) is analogous to   the equivalence of the capacity region of the multiple-access channel and the rate region achieved via time sharing and successive decoding, see~\cite[Chapter~15]{Cover}.
\end{remk}

We note that since $\alpha_0$ is a bounded variable in the interval $[0,1]$, a one-dimensional search algorithm can be used to find $\alpha_0^*$ assuming the optimal $\mathbf{D}^*$ is known for any given $\alpha_0$. In the following, we discuss the unimodality property of the sum rate w.r.t. ${\alpha}_0$ which enables the application of an efficient search algorithm, namely the GSS algorithm, to find the optimal $\alpha_0^*$ assuming that the optimal $\mathbf{D}^*$ is known.  
In particular, a unimodal function has only one optimal point in a given bounded interval~\cite{GSS}. Increasing $\alpha_0$ has two effects on the sum rate, namely the RF multiple-access time increases and the quantization distortion  also increases because the RF fronthaul time decreases. In other words, by increasing $\alpha_0$, the sum rate first increases owing to the increasing RF multiple-access time, but ultimately decreases due to the decrease of the fronthaul capacity and the resulting increase of the distortion. Hence, the sum rate is a unimodal function of $\alpha_0$ and has only one locally optimum point in the closed interval of $[0,1]$. Employing the GSS algorithm to find the optimal $\alpha_0^*$ (see Section IV.B for details), optimization problem (\ref{Eq:Local}) simplifies to finding the optimal $\mathbf{D}^*$ for a given $\alpha_0$ which is tackled in the following.

\subsubsection{Reformulation of the Objective Function} Note that for a given $\alpha_0$, the problem is still non-convex in $\mathbf{D}$ since the objective function of the \textit{maximization} problem is convex in $\mathbf{D}$ (instead of concave). To convexify the problem, in the following, we present a reformulation of (\ref{Eq:Local}) w.r.t. $\mathbf{D}$ using the following lemma.

\begin{lem}[\hspace{-0.3mm}\cite{lemma}]\label{LemmaD} For any matrix $\mathbf{X}\in \mathbb{C}^{J\times J}$ which satisfies $\mathbf{X} \succ \mathbf{0}$, the following equation holds
\begin{IEEEeqnarray}{lll}\label{Eq:Lemma}
\mathrm{log_2}|\mathbf{X}^{-1}| = \underset{\mathbf{Y} \succeq \mathbf{0}}{\mathrm{max}}\,\,\mathrm{log_2}|\mathbf{Y}|-\dfrac{1}{\ln(2)}\mathrm{Tr}(\mathbf{Y}\mathbf{X})+\dfrac{J}{\ln(2)},
\end{IEEEeqnarray}
where the optimal solution to the right-hand side of (\ref{Eq:Lemma}) is achieved when $\mathbf{Y^*}=\mathbf{X}^{-1}$.
\end{lem}

Defining $\mathbf{X}=\mathbf{D}+\sigma^2\mathbf{I}_{MN}$ and $\mathbf{Y}=\mathbf{B}$, where $\mathbf{B}$ is a new auxiliary optimization matrix, and applying Lemma~\ref{LemmaD} to $I\left(\mathbf{x},\hat{\mathbf{y}}\right)$ in (\ref{Eq:C_sum_Local}) and replacing the original constraints with the equivalent constraints from Lemma~\ref{LemmaA}, we reformulate optimization problem (\ref{Eq:Local}) as follows
\begin{IEEEeqnarray}{llll}\label{Eq:TOTAL}
 \underset{\mathbf{D}_m\succeq \mathbf{0}\,\,\forall m,}{\underset{\alpha_0\in[0,1] ,\mathbf{B}\succeq \mathbf{0}}{\mathrm{maximize}}}\,\,T=\alpha_0 W^{\mathrm{rf}}\Big[ \mathrm{log_2}|\mathbf{H}\boldsymbol{\Sigma}\mathbf{H}^{\mathsf{H}}+\mathbf{D} +\sigma^2\mathbf{I}_{MN}| \vspace{-4mm}\\
 \qquad\qquad\quad\,\,\,+\mathrm{log_2}|\mathbf{B}|-\dfrac{1}{\ln(2)}\mathrm{Tr}\left( \mathbf{B}\left(\mathbf{D}+\sigma^2\mathbf{I}_{MN}\right)\right)\Big] \nonumber\\
 \mathrm{subject\,\, to} \,\,\,  \mathrm{C}_{\mathcal{S}}:\,\,\alpha_0 f_s \sum_{\forall m\in \mathcal{S}}G_m(\mathcal{S})I\left(\mathbf{y}_m,\hat{\mathbf{y}}_m\right)\nonumber\\
\qquad\qquad\,\,\,\ \leq (1-\alpha_0)G(\mathcal{S}) +\sum_{\forall m\in \mathcal{S}}G_m(\mathcal{S}){C}_m^{\mathrm{fso}},\,\,\ \forall\mathcal{S} \subseteq \mathcal{M}.\nonumber
\end{IEEEeqnarray}
Although, for a given $\alpha_0$, optimization problem (\ref{Eq:TOTAL}) is still jointly non-convex in $(\mathbf{D},\mathbf{B})$, the problem is convex w.r.t. the individual variables. This allows the use of ACO to find a suboptimal solution of the problem in terms of $(\mathbf{D},\mathbf{B})$ for any given $\alpha_0$. In particular, for given $\mathbf{B}$, the problem is convex w.r.t. $\mathbf{D}$ and  for given $\mathbf{D}$, the problem is convex w.r.t. $\mathbf{B}$ and has the following optimal closed-form solution based on Lemma~\ref{LemmaD}
\begin{IEEEeqnarray}{lll}\label{Eq:B}
\mathbf{B}^*=\left(\mathbf{D}+\sigma^2\mathbf{I}_{MN}\right)^{-1}.
\end{IEEEeqnarray}
%

 In the following subsection, we propose a nested loop algorithm consisting of outer and inner loops by exploiting Lemma~\ref{LemmaA} and Lemma~\ref{LemmaD}, respectively.

\subsection{Proposed Algorithm}

Here, we propose an algorithm consisting of an outer loop, i.e., Algorithm~1, to find $\alpha_0$ based on GSS, and an inner loop, i.e., Algorithm~2, to find $(\mathbf{D},\mathbf{B})$ based on ACO, respectively~\cite{GSS}. In the following, we describe the outer and inner loops given by Algorithms~1~and~2, respectively, in detail.

\begin{figure}
\centering
\scalebox{0.58}{
\pstool[width=1.7\linewidth]{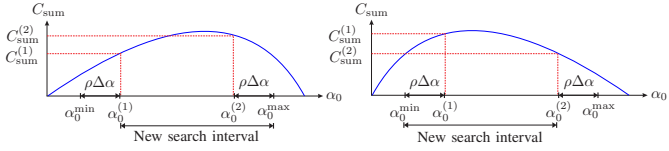}{
\psfrag{Csum}[c][c][1]{$C_{\mathrm{sum}}$}
\psfrag{al}[c][c][1]{$\alpha_0$}
\psfrag{amin}[c][c][1]{$\alpha_0^{\mathrm{min}}$}
\psfrag{amax}[c][c][1]{$\alpha_0^{\mathrm{max}}$}
\psfrag{a1}[c][c][1]{$\alpha_0^{(1)}$}
\psfrag{a2}[c][c][1]{$\alpha_0^{(2)}$}
\psfrag{da}[c][c][1]{$\rho\Delta\alpha$}
\psfrag{NewS}[c][c][1]{New search interval}
\psfrag{C1}[c][c][1]{$C_{\mathrm{sum}}^{(1)}$}
\psfrag{C2}[c][c][1]{$C_{\mathrm{sum}}^{(2)}$}
} } \vspace{-0.3cm}
\caption{Two possible cases for narrowing the search interval using GSS.}
\label{Fig:GSS}\vspace{-0.3cm}
\end{figure}

\textit{Outer Loop:}
In this loop, we employ the iterative GSS algorithm to maximize the sum rate w.r.t. $\alpha_0$. Suppose $[\alpha_0^{\mathrm{min}},\alpha_0^{\mathrm{max}}]$ is the search interval in a given iteration. Thereby, the GSS algorithm requires two evaluations of the sum rate in each iteration at the intermediate points $(\alpha_0^{(1)},\alpha_0^{(2)})$ using the inner loop, i.e., Algorithm~2. For GSS,  $(\alpha_0^{(1)},\alpha_0^{(2)})$ is obtained as 
\begin{IEEEeqnarray}{llll}\label{Eq:GSS_Points}
(\alpha_0^{(1)},\alpha_0^{(2)})=\left(\alpha_0^{\mathrm{min}}+\rho\Delta\alpha,\alpha_0^{\mathrm{max}}-\rho\Delta\alpha\right),
\end{IEEEeqnarray}
where $\Delta\alpha=\alpha_0^{\mathrm{max}}-\alpha_0^{\mathrm{min}}$ and $\rho=1-\dfrac{1}{\phi}$ with the so-called golden ratio $\phi=(1+\sqrt{5})/{2}\approx 1.61803$~\cite{GSS}. The GSS algorithm is schematically illustrated in Fig.~\ref{Fig:GSS}  and is given in Algorithm~1 for the problem considered in this paper. In Algorithm~1, $\epsilon$ is a small positive number which is used in line~10 for termination if the desired accuracy of the GSS algorithm in finding $\alpha_0^*$ is achieved. We note that the search interval is narrowed with a reduction factor of $1-\rho$ in each iteration.

\textit{Inner Loop:} 
In the inner loop, $\alpha_0$ is fixed as it was obtained in the outer loop. Here, we employ ACO to find a stationary point of the problem w.r.t. $\mathbf{B}$ and $\mathbf{D}$ for the fixed $\alpha_0$. The proposed ACO method is concisely given in Algorithm~2 where $\varepsilon$ is a small positive number used in the termination condition in line~7. We note that in the next subsection, we provide a modified ACO method which is easier to implement in popular numerical solvers such as CVX, which is also included in Algorithm~2 in blue italic font due to space limitations. 

Having $(\alpha_0^*,\mathbf{D}_1^*,\dots,\mathbf{D}_M^*)$, the RF time allocation variables for each fronthaul link, $\alpha_m$, are obtained as
\begin{IEEEeqnarray}{lll}\label{Eq:alpha_m}
\alpha_m^*=\frac{\alpha_0^*I\left(\mathbf{y}_m,\hat{\mathbf{y}}_m\right)\at[\big]{\mathbf{D}_m^*}-C_m^{\mathrm{fso}}}{C_m^{\mathrm{rf}}}, \,\,\forall m\in \mathcal{M}.
\end{IEEEeqnarray}
We note that Algorithms~1~and~2 find a stationary point of (\ref{Eq:TOTAL}) due to the non-convexity of the problem and the ACO employed in the inner loop.

\begin{remk}
We note that the proposed algorithm, i.e., the combination of Algorithms~1 and 2, finds in general a suboptimal solution of (\ref{Eq:Local}) due to the ACO used in the inner loop, i.e., Algorithm~2. 
\end{remk}

\subsection{Further Discussion}
 Although for given $\alpha_0$ and $\mathbf{B}$, optimization problem (\ref{Eq:TOTAL}) is convex, its implementation  using popular numerical solvers, such as CVX \cite{cvx}, can be challenging. More specifically, the implementation of line~5 of Algorithm~2 in CVX is not directly possible since the current version of CVX, i.e., CVX 2.1, does not have a built-in convex function that can directly handle $I\left(\mathbf{y}_m,\hat{\mathbf{y}}_m\right)$ in constraint $\mathrm{C}_{\mathcal{S}}$ in (\ref{Eq:TOTAL}) \cite{cvx}. In the following, we propose an equivalent reformulation of $\mathrm{C}_{\mathcal{S}}$ denoted by $\widetilde{\mathrm{C}}_{\mathcal{S}}$ to address this issue. Defining $\mathbf{X}=\left(\mathbf{H}_m\boldsymbol{\Sigma}\mathbf{H}_m^{\mathsf{H}}+\mathbf{D}_m+\sigma^2\mathbf{I}_{N}\right)^{-1}$ and $\mathbf{Y}=\mathbf{A}_m$, where $\mathbf{A}_m$ is a new auxiliary optimization matrix, and using (\ref{Eq:Lemma}) in Lemma~\ref{LemmaD}, we can upper bound $I\left(\mathbf{y}_m,\hat{\mathbf{y}}_m\right)$ by $R^{\mathrm{ub}}\left(\mathbf{D}_m,\mathbf{A}_{m}\right)$ as 
\begin{IEEEeqnarray}{lll}\label{Eq:R_ub}
I\left(  \mathbf{y}_m,\hat{\mathbf{y}}_m\right) \nonumber \\
=   -\mathrm{log_2}\left\vert \left(\mathbf{H}_m\boldsymbol{\Sigma}\mathbf{H}_m^{\mathsf{H}}+\mathbf{D}_m+\sigma^2\mathbf{I}_{N}\right)^{-1}\right\vert-\mathrm{log_2}\left\vert \mathbf{D}_m\right\vert\nonumber\\
\leq R^{\mathrm{ub}}\left(\mathbf{D}_m,\mathbf{A}_{m}\right)\nonumber\\=
-\mathrm{log_2}|\mathbf{A}_{m}|
+\dfrac{1}{\ln(2)}\mathrm{Tr}\left(\mathbf{A}_{m}\left(\mathbf{H}_m\boldsymbol{\Sigma}\mathbf{H}_m^{\mathsf{H}}+\mathbf{D}_m+\sigma^2\mathbf{I}_{N}\right)\right)\hspace{-0.1cm}\nonumber \\
-\dfrac{N}{\ln(2)}-\mathrm{log_2}\left\vert \mathbf{D}_m\right\vert,\quad \forall \mathbf{A}_{m}\succeq \mathbf{0}. 
\end{IEEEeqnarray}
Substituting the upper bound in (\ref{Eq:R_ub}) into the constraint $\mathrm{C}_{\mathcal{S}}$ in (\ref{Eq:TOTAL}), we obtain $\widetilde{\mathrm{C}}_{\mathcal{S}}$ which is convex in $\mathbf{D}$ if $\mathbf{A}_m$ is fixed and vice versa. It is not hard to see that constraint $\mathrm{C}_{\mathcal{S}}$ is always feasible when $\widetilde{\mathrm{C}}_{\mathcal{S}}$ is feasible and the two constraints are equivalent when
\begin{IEEEeqnarray}{lll}\label{Eq:A_m}
\mathbf{A}_{m}^{*}=\left(\mathbf{H}_m\boldsymbol{\Sigma}\mathbf{H}_m^{\mathsf{H}}+\mathbf{D}_m+\sigma^2\mathbf{I}_{N}\right)^{-1},\quad \forall m \in \mathcal{M}.
\end{IEEEeqnarray}
In other words, for a given $\mathbf{A}_{m}\neq \mathbf{A}_{m}^{*}$, the feasible set of $\widetilde{\mathrm{C}}_{\mathcal{S}}$ is strictly smaller than that of $\mathrm{C}_{\mathcal{S}}$. However, if we consider $\mathbf{A}_m\succeq \mathbf{0}$ as a new optimization variable which contains $\mathbf{A}_{m}=\mathbf{A}_{m}^{*}$ as a special case, the feasible sets of constraints $\widetilde{\mathrm{C}}_{\mathcal{S}}$ and $\mathrm{C}_{\mathcal{S}}$ become identical. Therefore, we rewrite the problem in (\ref{Eq:TOTAL}) as follows 
\begin{algorithm}[t] \label{Alg:Outer Loop}
\caption{GSS Algorithm (Outer Loop)}
\begin{algorithmic}[1] 
\STATE \textbf{input} $\epsilon$, $\alpha_0^{\mathrm{min}}=0$, and $\alpha_0^{\mathrm{max}}=1$.
\REPEAT 
\STATE Update $\alpha_0^{(1)}$ as in (\ref{Eq:GSS_Points}), obtain $\mathbf{D}^{(1)}$ from \textbf{{Algorithm~2}}, and compute $C_{\mathrm{sum}}^{(1)}$ from the objective function of (\ref{Eq:Local}).
\STATE Update $\alpha_0^{(2)}$ as in (\ref{Eq:GSS_Points}), obtain $\mathbf{D}^{(2)}$ from \textbf{{Algorithm~2}}, and compute $C_{\mathrm{sum}}^{(2)}$ from the objective function of (\ref{Eq:Local}).
\IF{$C_{\mathrm{sum}}^{(1)}\geq C_{\mathrm{sum}}^{(2)}$} 
\STATE Update $\alpha_0^{\mathrm{max}}=\alpha_0^{(2)}$,
\ELSIF{$C_{\mathrm{sum}}^{(1)}< C_{\mathrm{sum}}^{(2)}$} 
\STATE Update $\alpha_0^{\mathrm{min}}=\alpha_0^{(1)}$.
\ENDIF 
\UNTIL{$|\alpha_{0}^{\mathrm{max}}- \alpha_{0}^{\mathrm{min}}| \leq \epsilon$}
\STATE $\alpha_0^{*}\gets \dfrac{\alpha_{0}^{\mathrm{min}}+ \alpha_{0}^{\mathrm{max}}}{2}$  and obtain $\mathbf{D}^{*}$ from \textbf{Algorithm~2}.
\STATE \textbf{output} $\alpha_0^{*}$ and $\mathbf{D}^{*}$.
\end{algorithmic}
\end{algorithm}
\begin{algorithm}[t] \label{Alg:Inner Loop}
\caption{ACO (\textcolor{blue}{\textit{Modified ACO}}) Algorithm (Inner Loop)}
\begin{algorithmic}[1] 
\STATE \textbf{input} $\varepsilon$, $i=0$, $T^{\{0\}}=0$, ${\mathbf{D}}^{\{0\}}=d_0 \mathbf{I}$, and $\alpha_{0}$.
\REPEAT
\STATE $i \gets i+1$.
\STATE Given ${\mathbf{D}}^{\{i-1\}}$, update $\mathbf{B}^{\{i-1\}}$ (\textcolor{blue}{$\mathbf{\textit{B}}^{\textit{\{i\}}}$ \textit{and} $\mathbf{\textit{A}}_{\textit{m}}^{\textit{\{i\}}}$}) from the closed-form expression (\textcolor{blue}{\textit{expressions}})  given in (\ref{Eq:B}) (\textcolor{blue}{\textit{(\ref{Eq:B}) and (\ref{Eq:A_m}), respectively}}).
\STATE Given $\mathbf{B}^{\{i\}}$ and $\alpha_0$, update ${\mathbf{D}}^{\{i\}}$ by solving the convex problem (\ref{Eq:TOTAL}) (\textcolor{blue}{\textit{(\ref{Eq:TOTAL_CVX})}}).
\STATE Given $\alpha_0$, ${\mathbf{D}}^{\{i\}}$, and $\mathbf{B}^{\{i\}}$, compute $T^{\{i\}}$ from the objective function in (\ref{Eq:TOTAL}) (\textcolor{blue}{\textit{(\ref{Eq:TOTAL_CVX})}}).
\UNTIL{$|T^{\{i\}}-T^{\{i-1\}}|\leq \varepsilon$}
\STATE \textbf{output} $\mathbf{D}={\mathbf{D}}^{\{i\}}$.
\end{algorithmic}
\end{algorithm}
\begin{IEEEeqnarray}{llll}\label{Eq:TOTAL_CVX}
 \underset{\mathbf{B}\succeq \mathbf{0},\mathbf{A}_m\succeq \mathbf{0},\,\,\forall m}{\underset{\alpha_0\in[0,1] ,\mathbf{D}_m\succeq \mathbf{0},\,\,\forall m}{\mathrm{maximize}}}\,T=\alpha_0 W^{\mathrm{rf}}\Big[ \mathrm{log_2}|\mathbf{H}\boldsymbol{\Sigma}\mathbf{H}^{\mathsf{H}}+\mathbf{D} +\sigma^2\mathbf{I}_{MN}| \nonumber\vspace{-2mm}\\
 \qquad\qquad\qquad +\mathrm{log_2}|\mathbf{B}|-\dfrac{1}{\ln(2)}\mathrm{Tr}\left( \mathbf{B}\left(\mathbf{D}+\sigma^2\mathbf{I}_{MN}\right)\right)\Big] \quad\\
 \mathrm{subject\,\, to} \,\,\,  \widetilde{\mathrm{C}}_{\mathcal{S}}:\,\,\alpha_0 f_s \sum_{\forall m\in \mathcal{S}}G_m(\mathcal{S})R^{\mathrm{ub}}\left(\mathbf{D}_m,\mathbf{A}_{m}\right)\nonumber\\
\qquad\qquad\,\,\,\ \leq (1-\alpha_0)G(\mathcal{S}) +\sum_{\forall m\in \mathcal{S}}G_m(\mathcal{S}){C}_m^{\mathrm{fso}},\,\,\ \forall\mathcal{S} \subseteq \mathcal{M}.\nonumber
\end{IEEEeqnarray}
Solving (\ref{Eq:TOTAL_CVX}) w.r.t. $\alpha_0$, $\mathbf{D}$, $\mathbf{B}$, and $\mathbf{A}_m,\,\,\forall m$ is equivalent to solving (\ref{Eq:TOTAL}) w.r.t. $\alpha_0$, $\mathbf{D}$, and $\mathbf{B}$ which is in turn equivalent to solving the original problem in (\ref{Eq:Local}) w.r.t. $\boldsymbol{\alpha}$ and $\mathbf{D}$. The advantage of optimization problem (\ref{Eq:TOTAL_CVX}) is that existing numerical solvers can directly be employed to solve it. Based on (\ref{Eq:TOTAL_CVX}), a modified ACO algorithm w.r.t. $\mathbf{D}_m$, $\mathbf{B}$, and $\mathbf{A}_m,\,\forall m$ can be developed. The necessary changes are provided in blue italic font in Algorithm~2.

\section{Simulation Results}
In the following, we first present the simulation setup and the considered benchmark schemes. Subsequently, we provide simulation results to evaluate the performance of the proposed protocol compared to that of the benchmark schemes.

\subsection{Simulation Setup and Benchmark Schemes}
 We consider Rayleigh, Rician, and Gamma-Gamma fading for the RF multiple-access, RF fronthaul, and FSO channels, respectively~\cite{MyTCOM}. Moreover, we adopt the distance-dependent path loss models used in~\cite[Eq.~(5)]{MyTCOM}~and~\cite[Eq.~(2)]{MyTCOM} for the RF and FSO channels, respectively. Due to space constraints, we do not provide the equations for the path loss and fading models here and refer the readers to \cite{MyTCOM}. Unless stated otherwise, the values of the parameters for the RF and FSO links used in our simulations are given in Table~I where the noise power at the RF receivers is given by $[\sigma^2]_{\mathrm{dB}}=W^\mathrm{rf}N_{0}+N_{F}$, where $N_{0}$ and $N_{F}$ are defined in Table.~I. In particular, we generate random fading realizations for $10^3$ fading blocks and compute the sum rate of the proposed protocol for the solution found with Algorithms~1 and 2. Moreover, we assume  $K=8$, $M=2$, $N=8$, $L=64$, and $f_s=40$~MHz. In Algorithms~1~and~2, we use $\epsilon=0.02$ and $\varepsilon=0.01$ Mbps, respectively.
 
We consider the following two benchmark schemes. \textit{i) FSO-only fronthaul with scalar quantization (FSO-SQ):} For this benchmark scheme, we assume $\alpha_0=1$ and scalar quantization at the RUs where the signals at each RU's antennas are quantized independently with identical rates. 
 \textit{ii) FSO-only fronthaul with vector quantization (FSO-VQ):} Again, we assume $\alpha_0=1$; however, vector quantization is employed at the RUs. We note that although \cite{WeiYu_MIMO} does not consider FSO fronthaul links, the  optimization problem for fronthaul compression of FSO-VQ is similiar to that considered in~\cite{WeiYu_MIMO}.  
By comparing the proposed hybrid RF/FSO system to the above benchmark schemes, we are able to quantify how much performance gain can be obtained by vector quantization compared to scalar quantization and how much gain can be achieved by RF time allocation compared to the FSO-only schemes.

\begin{table}
\label{Table:Parameter}
\caption{Simulation Parameters~\cite{MyTCOM,FSO_Vahid}.\vspace{-0.2cm}} 
\begin{center}
\scalebox{0.6}
{
\begin{tabular}{|| c | c  | c ||}
  \hline
   \multicolumn{3}{||c||}{\textbf{RF Link}}\\ \hline \hline    
 Symbol & Definition & Value \\ \hline \hline
 $d^{\mathrm{ac}}$ & Distance between the MUs and the RUs & $100$ m \\ \hline
  $d^{\mathrm{fr}}$ & Distance between the RUs and the CU & $500$ m \\ \hline
   $d^{\mathrm{RF}}_{\mathrm{ref}}$ & Reference distance of the RF link & $5$ m \\ \hline
   $P_{k}$ & Transmit power of MU~$k$ & $16$ dBm \\ \hline 
   $\bar{P}_{m}$ & Transmit power of RU~$m$ & $33$ dBm \\ \hline  
 $(G^{\mathrm{Tx}}_{\mathrm{MU}}, G^{\mathrm{Rx}}_{\mathrm{RU}})$  & Antenna gains for the RF  multiple-access link & $(0,10)$ dBi \\ \hline
   $(G^{\mathrm{Tx}}_{\mathrm{RU}}, G^{\mathrm{Rx}}_{\mathrm{CU}})$   & Antenna gains for  RF fronthaul link & $(10,10)$ dBi \\ \hline
  $N_{0}$ & Noise power spectral density & $-114$ dBm/MHz \\ \hline 
  $N_{F}$ & Noise figure at the RF receivers & $5$ dB \\ \hline
   $\lambda^{\mathrm{RF}}$ & Wavelength of RF signal & $85.7$ mm  \\ \hline 
      $W^{\mathrm{RF}}$ & Bandwidth of RF signal & $40$ MHz \\ \hline 
     $\Omega$  & Rician fading factor & $6$ dB  \\ \hline  
    $\nu$ &  RF path-loss exponent & $3.5$ \\ \hline\hline 
       \multicolumn{3}{||c||}{\textbf{FSO Link}}\\ \hline \hline    
 Symbol & Definition & Value \\ \hline \hline
  ${P}^{\mathrm{fso}}_{m}$ & FSO transmit power of RU~$m$ & $13$ dBm \\ \hline 
  $\delta^2$ & Noise variance at the FSO receivers & $10^{-14}$ $\mathrm{A}^2$ \\ \hline 
   $\lambda^{\mathrm{FSO}}$ & Wavelength of FSO signal & $1550$ nm  \\ \hline
      $W^{\mathrm{FSO}}$ & Bandwidth of FSO signal & $1$ GHz  \\ \hline   
       $R$ & Responsivity of FSO photodetector & $0.5\frac{1}{\mathrm{V}}$   \\ \hline  
       $\phi$ & Laser divergence angle  & $2$ mrad  \\ \hline  
       $r$ & Aperture radius  & $10$ cm  \\ \hline      
              $(\Theta,\Phi)$ & Parameters of GGamma fading  & $(2.23,1.54)$  \\ \hline        
\end{tabular}
}
\end{center}
\vspace{-0.7cm}
\end{table}

\subsection{Performance Evaluation}

\begin{figure}
  \centering
\resizebox{1\linewidth}{!}{\psfragfig{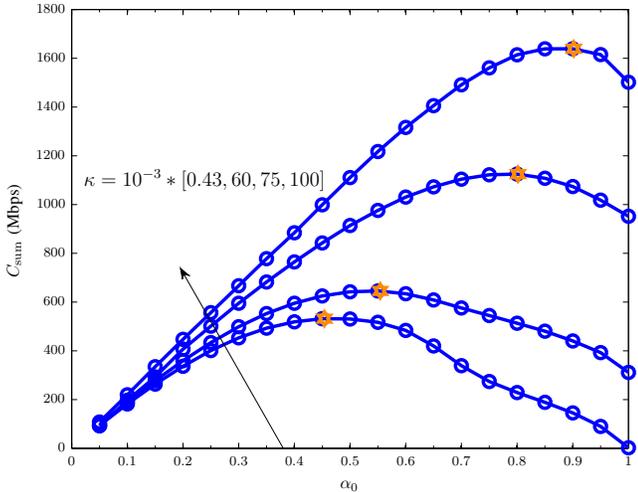}} \vspace{-1cm}
\caption{Average system sum rate (Mbps) vs. $\alpha_0$ for different values of the weather-dependent attenuation coefficient of the FSO fronthaul links,  $\kappa$. The value of $\kappa$ decreases along the direction of the arrow. Yellow star points are the optimal values of the sum rate which are found using Algorithm~1.} \vspace{-0.5cm}
\label{Fig:SumRate_Alpha}
\end{figure}

Fig.~\ref{Fig:SumRate_Alpha} shows the sum rate vs. $\alpha_0$ for the proposed protocol and different values of the weather-dependent attenuation coefficient of the FSO links $\kappa$. As can be observed, by increasing $\alpha_0$, the sum rate first increases owing to the increasing RF multiple-access time, but ultimately decreases due to the decrease of the fronthaul capacity. This shows the unimodality of the sum rate w.r.t. $\alpha_0$ which is discussed in Section~IV-B, i.e., each sum-rate curve in Fig.~\ref{Fig:SumRate_Alpha} has only one local optimum which is the global optimum too. Moreover, the optimal $\alpha_0^*$ (denoted by a yellow star in Fig.~\ref{Fig:SumRate_Alpha}) is found by Algorithm~1. Note that as the weather conditions deteriorate, i.e., as $\kappa$ increases, more RF time is allocated to the fronthaul links to compensate the loss in the quality of the FSO links, i.e., $\alpha_0^*$ decreases. Moreover, as expected, the sum rate decreases as weather conditions deteriorate, i.e., $\kappa$ increases.

In Fig.~\ref{Fig:SumRate_d_fr}, we show the sum rate of the proposed protocol and the benchmark schemes vs. the length of the fronthaul link, denoted by  $d^{\mathrm{fr}}$, i.e., the distance between the RUs and the CU, for different values of $\kappa$.
We observe from Fig.~\ref{Fig:SumRate_d_fr} that as $d^{\mathrm{fr}}$ increases, the sum rates of the proposed protocol and the benchmark schemes deteriorate due to the reduction of the fronthaul capacity. However, the performance degradation due to increasing $d^{\mathrm{fr}}$ is more severe for unfavourable weather conditions, i.e., for larger $\kappa$. In particular, for small $d^{\mathrm{fr}}$, the signal-to-noise-ratios (SNRs) of the FSO links are very high such that the capacities of the FSO links approach their maximum possible value for the adopted OOK modulation, i.e., $1$ Gbits/sec for the parameters considered here. However, as $d^{\mathrm{fr}}$ increases, beyond a certain value, the SNR is so low such that the capacities of the FSO links approach zero. The value of  $d^{\mathrm{fr}}$ above which the FSO links become unavailable depends on the weather conditions, e.g., for $\kappa=42\times 10^{-3}$ and $125\times 10^{-3}$, this distance is $950$~m and $400$~m, respectively. Hereby, for the proposed protocol, RF time allocation compensates the loss in the quality of the FSO links and a non-zero sum rate can still be achieved whereas the sum rate of the benchmark schemes, which do not have an RF fronthaul backup, drops to zero. For instance, from Fig.~\ref{Fig:SumRate_d_fr}, we observe that for heavy fog (i.e., $\kappa=125\times 10^{-3}$), although for $d^{\mathrm{fr}}=400$~m, the sum rate of the benchmark schemes is zero, the proposed protocol still achieves a sum rate of more than $500$~Mbps.
This clearly illustrates the benefits of having an RF fronthaul backup for ensuring a non-zero minimum achievable rate even under adverse atmospheric conditions. 
Finally, by comparing the sum rates of the FSO-VQ and FSO-SQ protocols in Fig.~\ref{Fig:SumRate_d_fr} when the FSO links are available, we can conclude that a considerable gain is achieved by vector quantization compared to scalar quantization due to the exploitation of the spatial correlation between the signals received at different antennas of each RU. We note that a similar performance gain was also reported in \cite{WeiYu_MIMO}.

\section{Conclusion}
In this paper, we considered sum-rate maximization for uplink C-RANs with hybrid RF/FSO fronthaul links. We optimized the RF time allocated to the multiple-access and  fronthaul links and the distortion matrices at the RUs. Since the resulting optimization problem was non-convex, we proposed a reformulation of the original problem which enabled the design of an efficient suboptimal algorithm based on GSS and ACO for solving the problem. Our simulation results revealed that a considerable gain can be achieved by the proposed protocol in comparison with benchmark schemes from the literature, especially when the FSO links experienced adverse atmospheric conditions.

\appendix

Due to space constraints, we only provide a sketch of the proof in the following. For notational simplicity, instead of considering the constraints in (\ref{Eq:TOTAL}), we consider constraints of the general form $a_m x_0\leq b_mx_m+c_m,\,\,\forall m\in\mathcal{M}$ with $x_m\geq 0$ and $\sum_{m=0}^M x_m = 1$ where $a_m$, $b_m$, and $c_m$ are positive constants. Dividing both sides of these constraints by $b_m$ and summing the right-hand sides and left-hand sides of all constraints with indices $m\in\mathcal{S}$, respectively, where $\mathcal{S}$ is a non-empty subset of $\mathcal{M}$, we obtain
\begin{IEEEeqnarray}{lll}
\sum_{m\in\mathcal{S}}\frac{a_m}{b_m} x_0\leq \sum_{m\in\mathcal{S}} x_m+ \sum_{m\in\mathcal{S}}\frac{c_m}{b_m}, \nonumber \\
x_0 \frac{ \sum_{m\in\mathcal{S}}\Big(a_m\prod_{m'\neq m, m'\in\mathcal{S}}b_{m'}\Big)}{ \prod_{m\in\mathcal{S}}b_m} \nonumber\\
\qquad\quad\leq \sum_{m\in\mathcal{S}} x_m + \frac{ \sum_{m\in\mathcal{S}}\Big(c_m\prod_{m'\neq m, m'\in\mathcal{S}}b_{m'}\Big)}{\prod_{m\in\mathcal{S}}b_m}, \nonumber \\
x_0 \sum_{m\in\mathcal{S}}a_m G_m(\mathcal{S}) \overset{(a)}{\leq} (1-x_0)G(\mathcal{S}) + \sum_{m\in\mathcal{S}}c_m G_m(\mathcal{S}),\quad
\end{IEEEeqnarray}
where for inequality $(a)$, we used definitions $G(\mathcal{S})=\prod_{m\in\mathcal{S}}b_m$ and $G_m(\mathcal{S})=\prod_{m'\neq m, m'\in\mathcal{S}}b_{m'}$ and the inequality $\sum_{m\in\mathcal{S}} x_m\leq 1-x_0$ which in general enlarges the corresponding  feasible set compared to that for the original constraints. However, the original feasible set defined by the constraint in (\ref{Eq:TOTAL}) for $\forall m\in\mathcal{M}$ is  identical to the feasible set of inequality $(a)$ if all $\mathcal{S}\subseteq \mathcal{M}$ are considered~\cite[Chapter~15]{Cover}. This completes the proof.

\begin{figure}
  \centering
\resizebox{1\linewidth}{!}{\psfragfig{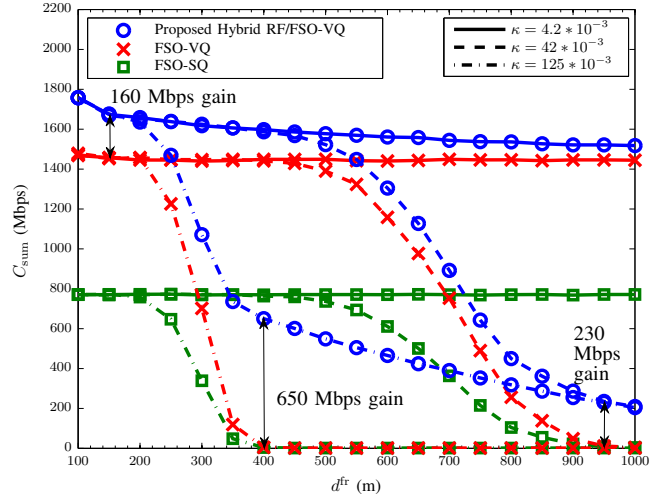}} \vspace{-1cm}
\caption{Average system sum rate (Mbps) vs. the length of the fronthaul links,  $d^{\mathrm{fr}}$, in meter for $\kappa=4.2 \times 10^{-3},42 \times 10^{-3},125\times 10^{-3}$ which correspond to  haze,  moderate fog, and heavy fog weather conditions, respectively~\cite{FSO_Vahid}.} \vspace{-0.5cm}
\label{Fig:SumRate_d_fr}
\end{figure}

\bibliographystyle{IEEEtran}
\bibliography{My_Citation_24-03-2017}

\end{document}